\documentclass[twocolumn,showpacs,amssymb,preprintnumbers,prc,floatfix]{revtex4}
\usepackage{epsfig,dcolumn}
\usepackage{graphicx,bm}

\begin{document}
\title{The isovector effective charge and the  staggering of the 2$^+$$\rightarrow$~$0^+$ transition
 probabilities in the Titanium isotopes}
\author{E. Caurier $^a$, F. Nowacki $^a$,
   A.~Poves $^b$}
\affiliation{
(a) IReS, B\^at27, IN2P3-CNRS/Universit\'e Louis
Pasteur BP 28, F-67037 Strasbourg Cedex 2, France\\
(b) Departamento de Fisica Te\'orica, C-XI. Universidad Aut\'onoma
de Madrid, E-28049, Madrid, Spain}
\date{\today}
\begin{abstract}
 In an effort to understand the magical  status of N=32 and N=34 at the
 very neutron rich edge, experiments have been carried out in the Titanium
 isotopes up to A=56. The measured staggering of the B(E2)'s is not reproduced by the
 shell model calculations using the best effective interactions. We argue
 that this may be related to the choice of the isovector effective
 charge and to the value of the N=34 neutron gap.

\end{abstract}
\pacs{21.10.Sf, 21.60.Cs, 23.20.Lv, 27.40.+z, 29.30.-h} 
\maketitle

 The two-body nucleon-nucleon interaction is different in the two isospin channels
 T=0 and T=1. Consequently, the mean field that it produces  varies depending
 on the relative number of neutrons and protons in a nucleus. At and around
 the stability line we find  the conventional ``magic'' numbers that,
 in the independent particle model of Mayer and Jensen \cite{Mayer}, were
 obtained by the addition of a strongly attractive spin-orbit potential to
 the isotropic harmonic oscillator. Approaching the proton drip line, these magic
 numbers seem to persist. The neutron drip line lies farther away of the stability
 valley, hence, the
 weight of the T=1 channel of the nuclear interaction relative to the T=0 channel
 increases in the mean field.
 One could wonder which would be the magic numbers of the  multi-neutron
 mean field, provided such a mean field made sense. One thing is clear; they will not 
 coincide with the standard ones.

 In the
 shell model context, this situation can be approached adding only neutrons -but many-
 to a well established doubly magic nucleus. In this case, the
 T=1 monopole interaction among the  valence neutrons  modifies the initial mean
 field felt by a single neutron on top of the doubly magic core. The important point
 to notice is that the eventual modifications of the shell structure should be due solely
 to the T=1 nucleon-nucleon interaction. One could rephrase that by saying that the magic numbers 
 for extremely neutron rich nuclei are dictated by the neutron-neutron interaction.
 The addition of valence protons activates the T=0 neutron-proton channel, leading to the 
 recovery of the known
 magic numbers. See in this respect the controversy in \cite{comment.zuker,reply.otsuka}
 on the monopole drift of the magic closures.

\begin{figure}[t]
  \begin{center}
    \leavevmode 
    \epsfig{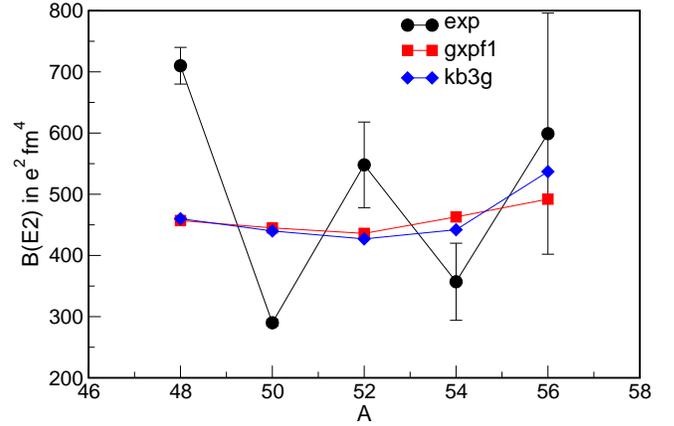}
     \caption{Theoretical B(E2)'s in the Titanium isotopes, calculated with effective charges  q$_{\pi}$=1.5~e 
              and q$_{\nu}$=0.5~e, compared with the experimental results}
    \label{fig:tibe2_A}
  \end{center}
\end{figure}

\begin{figure}[t]
  \begin{center}
    \leavevmode 
    \epsfig{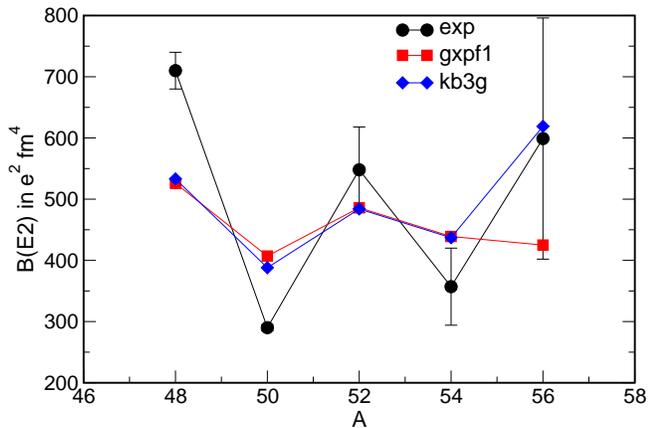}
     \caption{Theoretical B(E2)'s in the Titanium isotopes, calculated with effective charges  q$_{\pi}$=1.15~e 
              and q$_{\nu}$=0.8~e, compared with the experimental results}
    \label{fig:tibe2_B}
  \end{center}
\end{figure}

 A very handy study case is provided
 by the Calcium isotopes, that are nowadays known up to N=34. A strong sub-shell
 closure was shown  to exist at N=32  some time ago \cite{klotz}. The persistence of this sub-shell closure
 upon addition of protons has been the object  of many experimental studies recently \cite{janssens32}. As there is
 no spectroscopic information available on $^{54}$Ca,  less neutron rich, N=34, isotones  have  been
 explored \cite{liddick34} in order to check the magical status  of N=34.
 As a by-product of these studies, Dinca, Janssens {\it et al.} \cite{Dinca:2004} have
 made a comparison of the experimental excitation energies of the lowest 2$^+$ states
  and the  2$^+$~$\rightarrow$~$0^+$ transition probabilities in the even Titanium isotopes and the 
 predictions from the full $pf$ shell model calculations using the newly build effective interaction
 GXPF1 \cite{gxpf1}. This interaction produces  a large sub-shell gap at N=34, that disagrees with the
 experimental results \cite{fornal:2004}. An even newer interaction, GXPF1A~\cite{gxpf1a}, has been produced that
 predicts a less pronounced N=34 gap. However, none of them can explain the staggering of the 
 experimental B(E2)'s (see Fig.~2  in \cite{Dinca:2004} and Fig.~\ref{fig:tibe2_A}). This comes out as a
 surprise, the more so noticing that the interaction KB3G~\cite{kb3g:2001}, that does not predict
 a noticeable N=34 gap, is also unable to reproduce the trend of the data, even if, precisely at N=34's
 $^{56}$Ti, it behaves better,  as can be seen in Fig.~\ref{fig:tibe2_A}.

  Although there is general agreement on the use of an isoscalar effective  charge of +1.0~e in this region,
 the situation is less well established for the isovector effective charge. The calculations discussed above take it 
 equal to zero, while Dufour and Zuker~\cite{duzu} obtain +0.2~e and a very recent experimental
 and shell-model analysis of  the E2 transitions in the mirror pair  $^{51}$Fe-$^{51}$Mn \cite{Rietz:2004},
 using the interaction
 KB3G, concludes that the isovector effective charge is much larger; +0.65~e.  Dinca, Janssens {\it et al.} \cite{Dinca:2004}
 mention this new set of  effective charges but they conclude that ``although these values would induce a small 
 staggering in the calculated B(E2) values, they are not sufficient to bring experiment and theory into agreement''.
 However, a more detailed analysis turns out to be  worthwhile. In  Fig.~\ref{fig:tibe2_B} we show the results using 
 the new effective charges. The  comparison of   Fig.~\ref{fig:tibe2_A} and  Fig.~\ref{fig:tibe2_B} is very telling.
 The results with  KB3G and GXPF1 are identical for $^{48}$Ti, $^{50}$Ti, $^{52}$Ti, and $^{54}$Ti, irrespective
 of the set of effective charges used. Besides, using the new values of the effective charges, the experimental trend
 --and even the experimental values-- are now much better reproduced. Why is it so? Because a larger neutron effective charge
 tends to  amplify the contribution of the neutrons to the transition. If this contribution is large, 
 {\it i. e.} in absence of shell closure, it can override the effect of the reduction of the proton effective charge.
 If the neutrons are closed, the effect goes in the opposite direction. And
 this is what is seen in   Fig.~\ref{fig:tibe2_B}: in $^{48}$Ti, N=26, the B(E2) goes up, in $^{50}$Ti, N=28, 
 down, in $^{50}$Ti, N=30, up and finally, in $^{54}$Ti, N=32, down. The staggering is ready, and N=28 and N=32
 reaffirm their sub-shell status.

  We have deliberately left  $^{56}$Ti apart, because here, using the new isovector effective charge,
  the results of the two calculations diverge, reflecting their different underlying wave functions, in turn 
  dictated by their very different N=34 gaps. Consistently with the discussion above, the new effective charges 
  reduce the B(E2) value of GXPF1 because it closes the neutrons at N=34. On the contrary, they enhance 
  the KB3G value that does not have such a closure.
    
  In summary, the puzzling disagreement between theory and experiment in the lowest transitions of the Titanium isotopes
  can shed unexpected light into two apparently disconnected topics: The N=34 sub-shell closure far from stability and
  the value of the isovector effective charge for E2 transitions.

This work is partly supported by the IN2P3(France) CICyT(Spain)
collaboration agreements. AP's work is supported by  a grant of the DGI-MEC (Spain),
code BFM2003-1153.

\end{document}